\documentclass[conference]{IEEEtran}

\pdfoutput=1

\usepackage{graphicx}
\usepackage{amssymb}
\usepackage{amsmath}
\usepackage{url}
\usepackage{algorithm}
\usepackage[noend]{algorithmic}

\newcommand{\Spectra}{{\sc Spectra}}

\newcommand{\squishlist}{
 \begin{list}{$ullet$}
  { \setlength{\itemsep}{0pt}
     \setlength{\parsep}{3pt}
     \setlength{\topsep}{3pt}
     \setlength{\partopsep}{0pt}
     \setlength{\leftmargin}{1.5em}
     \setlength{\labelwidth}{1em}
     \setlength{\labelsep}{0.5em} } }
\newcommand{\squishend}{
  \end{list}  }

\begin{document}

\title{Estimating the pattern frequency spectrum\\inside the browser}

\author{
\IEEEauthorblockN{Matthijs van Leeuwen}
\IEEEauthorblockA{Department of Computer Science\\KU Leuven, Belgium\\Email: matthijs.vanleeuwen@cs.kuleuven.be}
\and
\IEEEauthorblockN{Antti Ukkonen}
\IEEEauthorblockA{Helsinki Institute for Information Technology HIIT\\Aalto University, Finland\\Email: antti.ukkonen@aalto.fi}
}

\maketitle

\begin{abstract}
We present a browser application for estimating the number of frequent
patterns, in particular itemsets, as well as the \emph{pattern frequency
spectrum}. The pattern frequency spectrum is defined as the function
that shows for every value of the frequency threshold $\sigma$ the
number of patterns that are frequent in a given dataset. Our demo
implements a recent algorithm proposed by the authors for finding the
spectrum. The application is 100\% JavaScript, and runs in all modern
browsers. We observe that modern JavaScript engines can deliver
performance that makes it viable to run non-trivial data analysis
algorithms in browser applications.

\end{abstract}


\section{Introduction}

Recent years have shown a trend of
implementing more and more complex
pieces of software
as {\em browser applications},
programs that run inside a browser.
Example of these are
the Google Docs\footnote{{\tt http://docs.google.com}}
suite of office applications,
WriteLatex\footnote{{\tt http://www.writelatex.com}},
a collaborative tool for preparing documents using \LaTeX,
or RStudio Server\footnote{{\tt http://www.rstudio.com}},
a rich graphical user interface to
the {\sf R} statistical computing environment.
Key enabling technologies
in this have been the HTML5 standard,
as well as the JavaScript
programming language
(based on ECMAScript \cite{ecma}).
In particular,
developments in the
performance of JavaScript engines,
such as Chrome V8\footnote{{\tt
https://developers.google.com/v8/}}
(used in Google Chrome)
and Spidermonkey\footnote{{\tt
https://developer.mozilla.org/en-US/docs/\\Mozilla/Projects/SpiderMonkey}}
(used in Mozilla Firefox),
together with the WebGL API \cite{webgl}
are making it possible
implement web applications that
need CPU/GPU performance,
such as graphics intensive games 
(see e.g.~the Cubeslam experiment\footnote{
{\tt http://www.chromeexperiments.com/detail/cube-slam/}} by Google)
or scientific visualisations
(e.g.~the Brain Tractography Viewer\footnote{
{\tt http://www.chromeexperiments.com/detail/\\
brain-surface-and-tractography-viewer/}}
by D.~Ginsburg and R.~Pienaar.).

Browser applications are appealing
because they are easy to distribute
in comparison to
``native'' software packages.
Nothing is installed on the users' computer,
and hence the same application can be used
from different computers around the world.
Once the application is updated
at the server,
the latest version is available to all users.

On the other hand,
prototypes of data mining and machine learning algorithms
that are proposed in literature
are only rarely packaged in a manner
that are easy to install and use
by others than the authors of the paper.
Even if source code is made public,
taking it into use may require
expertise or other effort
that a user of the algorithm does not have.
This is unfortunate as
it makes difficult to reproduce experiments,
and give proper exposure to
novel data analysis methods
among practitioners.
Browser applications, however,
are in general easy to install and take into use:
the user merely has to load a ``web page''.
They could thus be
a good approach to
demonstrate algorithm prototypes,
and even provide tools for users.

\begin{figure}[t]
\centering
\includegraphics[width=0.85\columnwidth]{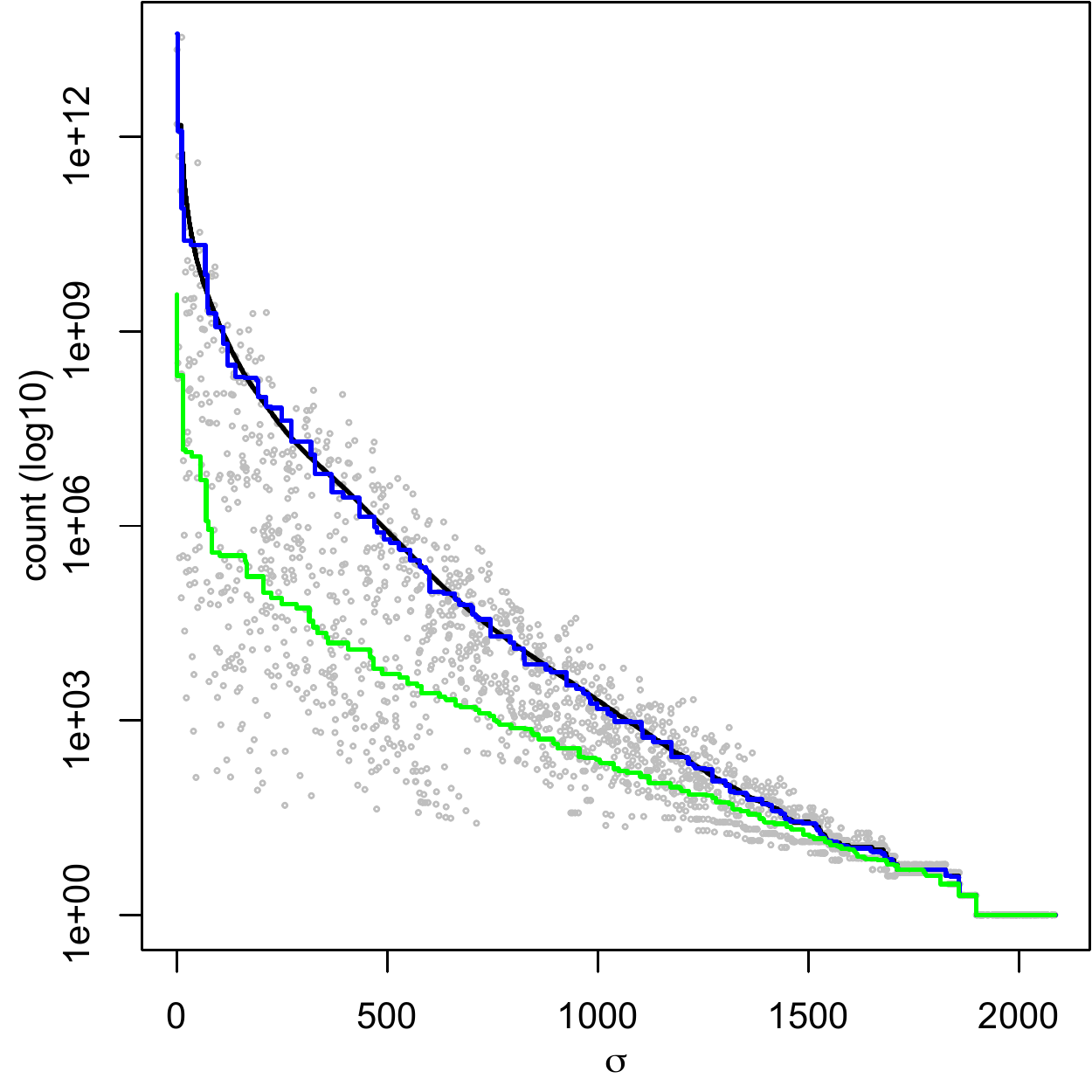}
\caption{(reproduced from \cite{spectra}) The {\em pattern frequency spectrum} (exact: black, estimate: blue) of the {\em Mammals} dataset. The estimate is computed by fitting an isotonic regression curve through the grey points. The green curve shows the spectrum of a random data having the same column marginals.}
\label{fig:intro}
\end{figure}
In this paper we showcase
a non-trivial data mining algorithm
implemented as a browser application
that carries out all computation
on the client computer.
Recently the authors proposed a method,
called \Spectra,
for estimating the {\em pattern frequency spectrum}
of a 0-1 dataset \cite{spectra}.
This is defined as
the curve that shows the
number of frequent patterns \cite{agrawal94apriori} in a dataset
as the function of a given frequency threshold $\sigma$
for all possible values of $\sigma$.
See Figure~\ref{fig:intro} for an example.
The same problem has also been considered by others,
but using a different technique \cite{BoleyG08,BoleyGG10}.

The spectrum can be used to determine
an appropriate value of $\sigma$
when mining patterns.
From Fig.~\ref{fig:intro} we can observe, for example, that
the number of patterns for $\sigma = 1000$
is in the order of one thousand,
but decreasing $\sigma$ to $500$
causes the number of patterns increase
by three orders of magnitude to one million.
However,
the spectrum
also serves as an interesting summary statistic of the database.
For instance,
by comparing the spectra of two datasets
we can assess to what extent
the datasets differ
in terms of the
joint distribution of the attributes.
In Fig.~\ref{fig:intro}
we have also computed the
spectrum of a random dataset that has the
same column marginals as the original dataset.
The resulting spectrum, shown in green,
is clearly different from the black/blue curves,
indicating that the original dataset
has structure not explained merely by the column marginals.

To compute the exact pattern frequency spectrum,
one must enumerate
(or at least count)
all patterns that are frequent for
$\sigma = 1$,
and output their respective frequencies.
For many datasets this is
either very slow,
or even impossible due to the
explosion in the number of frequent patterns
for small frequency thresholds.
The problem of counting frequent patterns
is in fact \#P-complete \cite{GunopulosKMSTS03}.
The \Spectra\ algorithm \cite{spectra}
is a fast method for computing
good estimates of the pattern frequency spectrum.
It is based on a modification to Knuth's algorithm
for estimating sizes of search trees \cite{knuth1975estimating},
as well as isotonic regression \cite{isoreg}.

The purpose of this paper is thus twofold.
On the one hand we want to {\em showcase the
\Spectra\ algorithm} inside a browser application,
and provide the community with
an easily accessible tool
for estimating frequent pattern counts.
On the other hand,
we want to argue that
{\em modern JavaScript engines
embedded in contemporary web browsers
can be
a viable platform for running
non-trivial data mining and machine learning algorithms}
that often have high performance requirements.
The application can be viewed at
\begin{quote}
{\tt http://anttiukkonen.com/spectra}
\end{quote}
and is best run in recent versions of Chrome or Safari,
although all reasonably modern browsers
should be supported.

\section{The \Spectra{} Algorithm}

In this section
we give a brief explanation of the
\Spectra\ algorithm.
Please see \cite{spectra}
for a full description as well as experimental results.

Knuth's algorithm \cite{knuth1975estimating}
for estimating the size of a combinatorial search tree
is based on the idea of
sampling random paths from the root of the tree to a leaf.
Let $\mathbf{d} = (d_0, \ldots, d_{h-1})$
denote the sequence of branching factors observed
along a path from the root (on level $0$) to a leaf (on level $h$).
The estimate 
produced by $\mathbf{d}$,
which we call the {\em path estimate},
is defined as
\begin{equation}
\label{eq:pathest}
\hat e(\mathbf{d}) = 1 + \sum_{i=1}^{h} \prod_{i=0}^{i-1} d_i.
\end{equation}
Observe that
if the tree were perfectly regular,
meaning that all nodes
that belong to the same level of the tree
had the same branching factor,
$\hat e(\mathbf{d})$
would be equal to the size of the tree
for any path.
Since search trees are rarely regular in practice,
Knuth's algorithm
{\em samples a number of different paths},
and uses the average of the $\hat e(\mathbf{d})$ values
as an estimate of the size of the tree.
It can be shown
that this is an unbiased estimate of
the tree size,
but the estimator has a high variance
\cite{knuth1975estimating}.

To use this method
for estimating the number of frequent patterns,
in particular itemsets,
we must consider that the patterns
do not form a tree,
but a {\em subset lattice}.
The set of patterns
that are frequent for a given threshold $\sigma$,
are located in this lattice
in the region between $\emptyset$
and the {\em positive border} \cite{GunopulosKMSTS03},
defined as the set of all {\em maximal} patterns.
A pattern is maximal
if it cannot be made more specific
without causing its frequency to drop below $\sigma$.

We estimate the number of frequent patterns for a given $\sigma$ by
sampling a number of paths
from the root of the lattice
up to (and including) the positive border,
compute the path estimates,
and take the average of these.
In case of itemsets,
sampling a path
corresponds to constructing an itemset
step-by-step
by adding one new item at every level of the lattice.
At every step the branching factor is
given by the number of extensions to
the current pattern
that are still frequent.
The item to be added
is sampled uniformly at random
from the set of possible extensions.

However, we need a small
modification to Eq.~\ref{eq:pathest}.
We must take into account that
a given node of the lattice
can be reached via multiple paths,
while in a tree the paths are unique.
As discussed in \cite{spectra},
it is sufficient to replace
Eq.~\ref{eq:pathest}
with the following:
\begin{equation}
\label{eq:pathest2}
e(\mathbf{d}) = 1 + \sum_{i=1}^{h} \frac{1}{i!} \prod_{i=0}^{i-1} d_i,
\end{equation}
where we have included
the normalisation term $1/i!$
to account for the $i!$
possible paths that can reach every node
on the $i$:th level.

This way we can estimate the
number of frequent patterns for
{\em one} frequency threshold $\sigma$.
Computing the pattern frequency spectrum
requires us to estimate
the number of patterns
for {\em all} meaningful values of $\sigma$.
A simple approach would be to
compute the estimates
for a number of fixed values of $\sigma$,
and use this as a coarse representation
of the frequency spectrum.
However,
turns out that we can
make more efficient use of the paths
by a procedure where we
use a {\em different} $\sigma$
for each sampled path,
and obtain a more accurate estimate.
Each $\sigma$
is drawn uniformly at random from
a specified interval.
We obtain thus a set of $N$ points
$P = \{(\sigma_1, \hat e_1), (\sigma_2, \hat e_2), \ldots, (\sigma_N, \hat e_N)\}$,
where $\sigma_k$ is the threshold for path $k$,
and $\hat e_k$ the corresponding path estimate
computed with Eq.~\ref{eq:pathest2}.
Then, rather than taking the average of the path estimates,
we fit an {\em isotonic regression} \cite{isoreg} curve
through the points in $P$.
This is a least squares fit to $P$
subject to the constraint that
the resulting curve must be decreasing.
The number of patterns
can only decrease as $\sigma$ grows,
hence we must constrain the
estimator in the same way.
In Figure~\ref{fig:intro}
the set $P$ is shown by the gray points,
and the corresponding isotonic least squares fit
is shown in blue
together with the exact frequency spectrum (black).


\section{Implementation}

The application consists of a 100\% JavaScript
implementation of the \Spectra\ algorithm.
Nothing is being sent to a server,
all processing is carried out locally at the client.
We have tried to
make the implementation as efficient as possible.
In particular,
we have followed programming guidelines
that would result in good performance
on the V8 JavaScript engine (Google Chrome) \cite{wilson2012}.

The algorithm spends most of the time
computing the branching factor at every step.
This involves finding all
extensions of the current pattern that are still frequent
(given $\sigma$).
For maximum performance,
the implementation uses a vertical data representation,
where every attribute is a bit vector.
Our implementation maintains a list of row identifiers
that support the current pattern,
and intersects this with
the column vectors of all possible extensions.
Counting the support is
done by scanning the resulting bit vectors
and using a lookup table to
quickly find the number of bits set to one
in every 16 bit segment.
Using bit vectors may become a bottleneck
for very large and sparse datasets
that have a large number of transactions and attributes.
However,
in our initial tests,
we found bit vectors to usually outperform
a variant that computes intersections of sorted lists of integers.

To find the isotonic regression curve,
we implemented the Pool Adjacent Violators (PAVA)
algorithm \cite{barlow1972statistical}.
This algorithm is straightforward to implement,
and is in practice very fast.
In comparison to sampling the paths
running PAVA is negligible.

\section{Performance evaluation}

\begin{table}[t]
\centering
\caption{Running times on different browsers (in sec)}
\label{table:res}
\begin{tabular}{l|rr|rrr}
\hline
\hline
dataset   & rows  & attrs. & Chrome & Firefox & Safari \\
\hline
Accidents & 340183 & 468    & 444    & 625     & 706    \\
Adult     & 48842  & 97     &  23    & 25      & 15     \\
Chess     & 3196   & 75     & 3.4    &  6.6    & 2.5    \\
Connect   & 67557  & 129    & 135    &  202    & 235    \\
Letrecog  & 20000  & 102    & 13     & 12     & 6.3    \\
Mammals   & 2183   & 121    & 1.2    &  1.9    & 0.9    \\
Mushroom  & 8124   & 119    & 4.6    & 6.7     & 3.5    \\
Pumsbstar & 49046  & 2088   & 117    &  145    & 83     \\
Waveform  & 5000   & 101    &  2.2   & 3.2     & 1.5    \\
\hline
\hline
\end{tabular}
\end{table}
We studied the performance of
the \Spectra{} implementation on different
JavaScript engines
by running it in a number of popular browsers.
We consider OSX versions of
Chrome 35.0 (V8),
Firefox 30.0 (SpiderMonkey), and
Safari 6.1.4 (Webkit)
on a 1.8GHz Intel i5 CPU.
We used publicly available datasets
with their properties shown in Table~\ref{table:res}.
In every case we used the algorithm to
estimate the frequency spectrum
by using 5000 random paths,
with every $\sigma$ drawn uniformly at random
from the interval $[1,1000]$.
We note that for practical applications
a smaller number of samples
can be sufficient.
Also, this only estimates the frequency spectrum
in for relatively ``small'' values of $\sigma$.
For larger $\sigma$ even
exact counting (not implemented in our demo)
can be a viable option
and hence estimation should
in practice
focus on the $[1,1000]$
range of the pattern frequency spectrum.

Table~\ref{table:res} shows the running times (in seconds)
for all combinations of dataset and browser.
There are substantial differences between
the different JavaScript engines,
with Webkit (Safari) having the best performance
except for the larger datasets (Accidents, Connect).
For the smaller datasets Safari is
about twice as fast as Firefox,
while the V8 engine of Chrome
overtakes the others by a clear margin
with Accidents and Connect.

We do not want to speculate
why the engines behave the way
we observe in Table~\ref{table:res}.
All engines are just-in-time compilers,
and e.g.~the heuristics for
deciding which functions and when to compile
may differ.
Also, this simple benchmark
is not to be taken as
a representative study of JavaScript engine performance.
It merely illustrates
their effect on the running time of \Spectra.

The running times of all engines,
however,
compare favourably to the
unoptimized C++ implementation
of \Spectra\ used in \cite{spectra}
by being often in fact faster.
For example,
using the JavaScript implementation
to estimate the number of patterns at $\sigma=7500$ in Pumsbstar
with 1000 paths
takes about 15 seconds with Firefox
(8.5 with Safari)
while the C++ implementation required 320 seconds
(Table 1 in \cite{spectra}).
Similar speedups are observed with other datasets.
We believe this to be due to
the horizontal data representation,
as well as other
design choices that were not motivated by performance
used in the C++ implementation.

\section{The Application}

\begin{figure*}[t]
\centering
\includegraphics[width=0.92\textwidth]{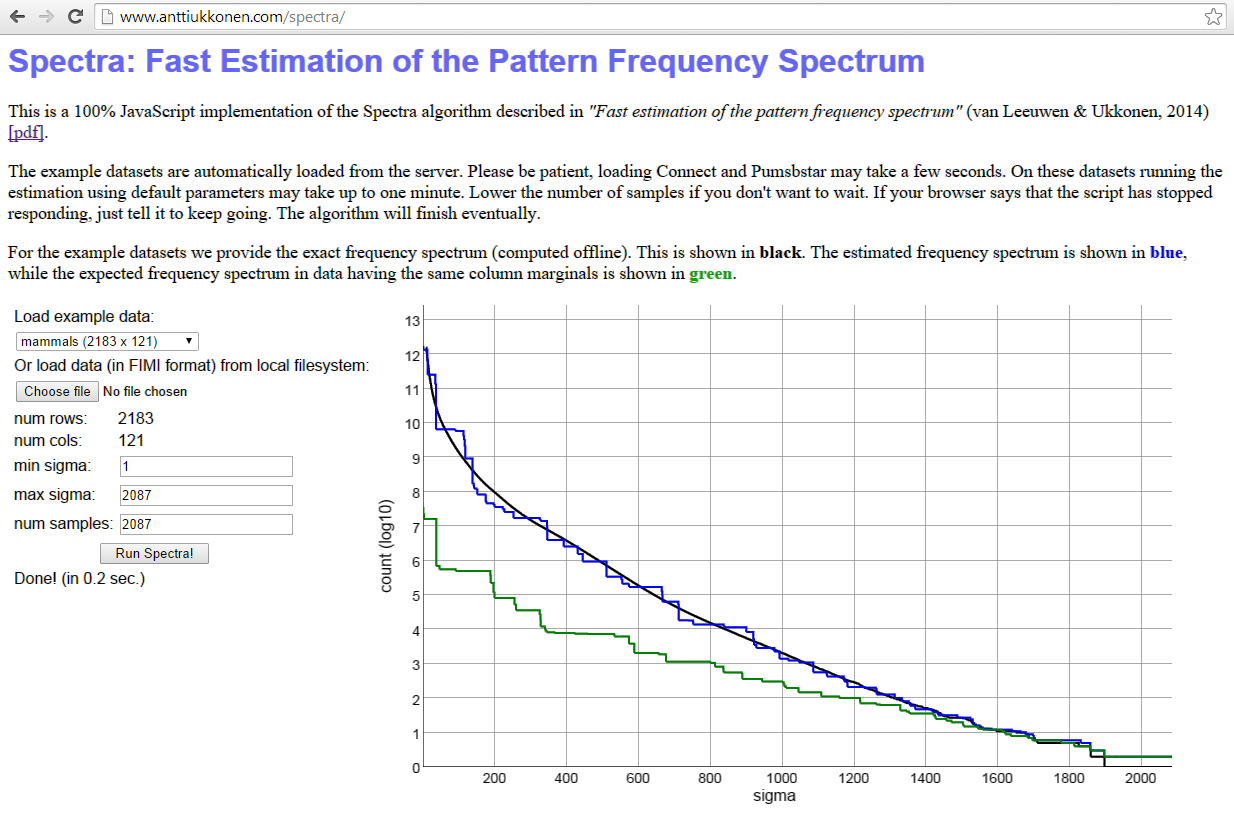}
\caption{Screenshot of the application running in Chrome. On the left we see that one of the predefined datasets (Mammals) has been loaded. The algorithm has estimated the pattern frequency spectrum for $\sigma \in [1,2087]$ using 1000 paths. The result is visualised on the right (blue curve), together with the exact frequency spectrum (in black, precomputed, loaded from the server), and the estimate for a frequency spectrum for a random dataset having the same column marginals as Mammals. The algorithm took 0.7 seconds to estimate the pattern frequency spectrum.}
\label{fig:screenshot}
\end{figure*}

The application contains the following basic functionalities:
\begin{enumerate}
\item Loading a user-supplied dataset in FIMI format\footnote{This
is the same format used by almost all public itemset mining tools. The input is a simple text file with one transaction on each row, where every transaction is a space-separated list of integers from the range $[1,m]$.}.
\item A list of benchmark datasets for which pre-computed exact frequency spectra are provided,  so that the estimates can be compared against these.
\item Computing the pattern frequency spectrum
in a given frequency threshold interval
using a given number of samples.
\item Computing the pattern frequency spectrum
of a random dataset having the same column marginals as the loaded dataset.
\item A visualisation of the resulting
pattern frequency spectra.
\end{enumerate}

Figure~\ref{fig:screenshot} shows
a screenshot of the application.
On the left is a panel where the user can
select a predefined dataset from a drop-down menu,
load a dataset in FIMI format from the local filesystem,
and modify different parameters of the algorithm.
Once a dataset has been loaded,
the parameters are automatically assigned default values.
These can be freely modified
before starting the algorithm.
Once the algorithm finishes running,
the visualisation on the right displays
both the
frequency spectrum estimated from the currently loaded dataset,
as well as an estimate of the frequency spectrum for
a dataset that has the same column marginals,
but is otherwise random.
If one of the predefined datasets was used,
the visualisation also shows
the exact precomputed
pattern frequency spectrum
that is loaded from the server
together with the data.

\section{Conclusion}

We presented a browser application for estimating the number of frequent
patterns, in particular itemsets, as well as the pattern frequency spectrum.

Using proprietary technologies such as
Java applets (Oracle) or Flash (Adobe),
it has been possible
for almost 20 years
to embed complex applications into web pages.
However,
these technologies have suffered
from interoperability and performance issues.
Recent developments in modern web browsers
have solved many of these problems
by converging to standards
that work across browsers
and platforms:
the same codebase can implement
an application that runs both on
a desktop and a mobile phone.

Our implementation of the \Spectra{} algorithm, for computing the pattern frequency spectrum, demonstrates that modern JavaScript engines can deliver performance that makes it viable to run non-trivial data analysis algorithms in browser applications.


\bibliographystyle{IEEEtran}
\bibliography{references}

\begin{thebibliography}{10}
\providecommand{\url}[1]{#1}
\csname url@samestyle\endcsname
\providecommand{\newblock}{\relax}
\providecommand{\bibinfo}[2]{#2}
\providecommand{\BIBentrySTDinterwordspacing}{\spaceskip=0pt\relax}
\providecommand{\BIBentryALTinterwordstretchfactor}{4}
\providecommand{\BIBentryALTinterwordspacing}{\spaceskip=\fontdimen2\font plus
\BIBentryALTinterwordstretchfactor\fontdimen3\font minus
  \fontdimen4\font\relax}
\providecommand{\BIBforeignlanguage}[2]{{%
\expandafter\ifx\csname l@#1\endcsname\relax
\typeout{** WARNING: IEEEtran.bst: No hyphenation pattern has been}%
\typeout{** loaded for the language `#1'. Using the pattern for}%
\typeout{** the default language instead.}%
\else
\language=\csname l@#1\endcsname
\fi
#2}}
\providecommand{\BIBdecl}{\relax}
\BIBdecl

\bibitem{ecma}
``{ECMAScript Language Specification (Standard ECMA-262)},'' 2011.

\bibitem{webgl}
\BIBentryALTinterwordspacing
{Khronos WebGL Working Group}, ``Webgl specification.'' [Online]. Available:
  \url{https://www.khronos.org/registry/webgl/specs/1.0/}
\BIBentrySTDinterwordspacing

\bibitem{spectra}
M.~van Leeuwen and A.~Ukkonen, ``Fast estimation of the pattern frequency
  spectrum,'' in \emph{Proceedings of ECML PKDD'14}, 2014, pp. 114--129.

\bibitem{agrawal94apriori}
R.~Agrawal and R.~Srikant, ``Fast algorithms for mining association rules in
  large databases,'' in \emph{Proceedings of VLDB'94}, 1994, pp. 487--499.

\bibitem{BoleyG08}
M.~Boley and H.~Grosskreutz, ``A randomized approach for approximating the
  number of frequent sets,'' in \emph{Proceedings of ICDM'08}, 2008, pp.
  43--52.

\bibitem{BoleyGG10}
M.~Boley, T.~G{\"a}rtner, and H.~Grosskreutz, ``Formal concept sampling for
  counting and threshold-free local pattern mining,'' in \emph{Proc. of
  SDM'10}, 2010, pp. 177--188.

\bibitem{GunopulosKMSTS03}
D.~Gunopulos, R.~Khardon, H.~Mannila, S.~Saluja, H.~Toivonen, and R.~Sharm,
  ``Discovering all most specific sentences,'' \emph{ACM Trans. Database
  Syst.}, vol.~28, no.~2, pp. 140--174, 2003.

\bibitem{knuth1975estimating}
D.~Knuth, ``Estimating the efficiency of backtrack programs,''
  \emph{Mathematics of computation}, vol.~29, no. 129, pp. 122--136, 1975.

\bibitem{isoreg}
R.~Barlow and H.~Brunk, ``The isotonic regression problem and its dual,''
  \emph{Journal of the American Statistical Association}, vol.~67, no. 337, pp.
  140--147, 1972.

\bibitem{wilson2012}
C.~Wilson, ``Performance tips for javascript in v8,''
  \url{http://www.html5rocks.com/en/tutorials/speed/v8/}, 2012.

\bibitem{barlow1972statistical}
R.~E. Barlow, D.~J. Bartholomew, J.~Bremner, and H.~D. Brunk, \emph{Statistical
  inference under order restrictions: the theory and application of isotonic
  regression}.\hskip 1em plus 0.5em minus 0.4em\relax Wiley New York, 1972.

\end{thebibliography}

\end{document}